# STOCHASTIC THEORY

# OF RELATIVISTIC QUANTUM MECHANICS

Maurice GODART,
Email = maurice.godart@skynet.be

## 0. Introduction

We have presented in a previous paper (arXiv:1206.2917v2[quant-ph]) arguments supporting the idea that the stochastic theory of quantum mechanics is a promising substitute to the orthodox theory in the non-relativistic domain. The questions now inevitably arise to know if it is possible to extend it to the relativistic domain and how to do it. In order to propose solutions to those problems we first transpose to the special context of the 4-dimensional Minkowski space all the definitions, equations and relations that are known to be valid for the Markov processes in the completely abstract mathematical context of an m-dimensional space. We shall therefore utilize methods not developed here and not assumed to be familiar to the average reader. Hence, we simply refer to the relevant literature. We then explain how the principles of relativity come into play. Along the way, we introduce ad hoc assumptions for the sole reason that they facilitate or even make possible the further development of the theory. This appears clearly in the choice of the specific expressions proposed for the drift vectors and the diffusion tensor. These hypotheses and the so-called Nelson equations are used to determine the diffusion tensor and the drift vectors characteristic of such a process. It is then possible to write down the Kolmogorov equations that are used to determine the conditional probability density, as well as the Fokker-Planck equation that can be used to determine the normal probability density. We describe the logical sequence of steps that leads to the solution of specific problems and we illustrate the resulting method by applying it to the special case of the free particle. We also suggest a possible correspondence between the relativistic and non-relativistic versions of the normal and conditional probability densities.

## 1. The relativistic Markov process

The co-ordinates of Minkowski space will be referenced by indices ranging from 0 to 3 and correspond to the time t and to the co-ordinates of the ordinary space according to the definitions:

$$\begin{cases} x^0 = ct \\ x^1 = x \\ x^2 = y \\ x^3 = z \end{cases} \tag{1-1}$$

The metric tensors $g_{ij}$ and $g^{ij}$ are defined by:

$$\left(g_{ij}\right) = \left(g^{ij}\right) = \begin{pmatrix} 1 & 0 & 0 & 0 \\ 0 & -1 & 0 & 0 \\ 0 & 0 & -1 & 0 \\ 0 & 0 & 0 & -1 \end{pmatrix} \tag{1-2}$$

and will be used in the usual way for raising or lowering freely covariant or contravariant indices of vectors and of tensors.

The basic hypothesis of the stochastic theory of relativistic quantum mechanics is that the trajectories of particles in the Minkowski space correspond to the sample functions of a diffusion process $\overline{x}$ whose components $x^i$ are expressed as random variables depending on a dynamical parameter. In classical relativistic mechanics, it is sometimes confusing but often convenient to use the proper time t of an observer as the dynamical parameter. This is no longer possible in relativistic stochastic theory where a clear distinction must be made between the random variable $x^0$ of the stochastic process and its dynamical parameter. This is why the time t that appears in the component $x^0$ of this stochastic process must be considered as a random variable $t(s,\omega)$ depending on a dynamical parameter that we shall denote subsequently by the different letter s in order to avoid any possible confusion.

The various definitions, equations and relations pertinent to the Markov and diffusion processes will be presented briefly. We shall always simplify their writing by adopting the summation convention of Einstein extended to the symbols $\partial_i \equiv \partial / \partial x^i$ and $\partial^2_{ij} \equiv \partial^2 / \partial x^i \partial x^j$.

We shall say that a stochastic process $x(t)$ is a Markov process if the condition:

$$p\left(\overline{x}_n, s_n \mid \overline{x}_0, s_0; \cdots ; \overline{x}_{n-1}, s_{n-1}\right) = p\left(\overline{x}_n, s_n \mid \overline{x}_{n-1}, s_{n-1}\right) \tag{1-3}$$

is verified for any increasing sequence $s_0 < s_1 < \cdots < s_{n-1} < s_n$ of values for the parameter s. Let us remember that the notation $p(*|*)$ designates the conditional probability density of the variables at the left of the vertical bar with respect to the variables at the right of it.

In the case where $n = 2$ the completely general rule:

$$p\left(\overline{x}_2, s_2 \mid \overline{x}_0, s_0\right) = \int_{-\infty}^{\infty} p\left(\overline{x}_2, s_2 \mid \overline{x}_0, s_0; \overline{x}_1, s_1\right) p\left(\overline{x}_1, s_1 \mid \overline{x}_0, s_0\right) d\overline{x}_1 \tag{1-4}$$

takes the particular form:

$$p\left(\overline{x}_2, s_2 \mid \overline{x}_0, s_0\right) = \int_{-\infty}^{\infty} p\left(\overline{x}_2, s_2 \mid \overline{x}_1, s_1\right) p\left(\overline{x}_1, s_1 \mid \overline{x}_0, s_0\right) d\overline{x}_1 \tag{1-5}$$

also called the Chapman-Kolmogorov equation. Note that the integrals here above are 4-fold integrals extending over the complete 4-dimensional Minkowski space

At first sight, the condition imposed to the sequence of the values of the parameter s to be increasing lets inevitably presage that a Markov process must be irreversible, but this is not true, because quite unexpectedly, it can be proved that the condition (1-3) is also verified for any decreasing sequence $s_0 > s_1 > \cdots > s_{n-1} > s_n$ of the parameter s. The property of reversibility of the Markov processes will often oblige us to duplicate definitions, equations and relations given hereafter. We shall present them a first time in the so-called forward case of an increasing sequence of values of the parameter s and we shall use the index + to remember this context. We shall present them a second time in the so-called backward case of a decreasing sequence of values of the parameter s and we shall use the index - to remember this context.

We shall study in more details the so-called diffusion processes that are actually Markov processes characterized by the fact that they verify the conditions:

$$\begin{cases} \int_{-\infty}^{\infty} \left(x_1^i - x_0^i\right) p\left(\bar{x}_1, s_1 \mid \bar{x}_0, s_0\right) d\,\bar{x}_1 = v_{\pm}^i\left(\bar{x}_0, s_0\right)\left(s_1 - s_0\right) + o^i\left(s_1 - s_0\right) \\ \int_{-\infty}^{\infty} \left(x_1^i - x_0^i\right)\left(x_1^j - x_0^j\right) p\left(\bar{x}_1, s_1 \mid \bar{x}_0, s_0\right) d\,\bar{x}_1 = 2\, w_{\pm}^{ij}\left(\bar{x}_0, s_0\right) \left| s_1 - s_0 \right| \\ \qquad\qquad + o^{ij}\left(s_1 - s_0\right) \\ \int_{-\infty}^{\infty} \left| \bar{x}_1 - \bar{x}_0 \right|^n p\left(\bar{x}_1, s_1 \mid \bar{x}_0, s_0\right) d\,\bar{x}_1 = o\left(s_1 - s_0\right) \quad \text{when} \quad n \geq 3 \end{cases} \tag{1-6}$$

depending on the sign of $s_1 - s_0$, where the symbols $o\left(s_1 - s_0\right)$ denote quantities that converge to 0 faster than $s_1 - s_0$ and where $\left| \bar{x}_1 - \bar{x}_0 \right|$ is a distance (in the mathematical sense) between the points $\bar{x}_1$ and $\bar{x}_0$. According to the classical terminology, the contravariant vectors $v_{\pm}^i\left(\bar{x}, s\right)$ are called the drift vectors and the contravariant tensors $w_{\pm}^{ij}\left(\bar{x}, s\right)$ are called the diffusion tensors.

If we introduce the conditional mathematical expectation:

$$E\left\{ f\left(\bar{x}_1\right) \mid \bar{x}_0 \right\} = \int f\left(\bar{x}_1\right) p\left(\bar{x}_1 \mid \bar{x}_0\right) d\bar{x}_1 \tag{1-7}$$

we can present the definitions of the drift vectors and of the diffusion tensors in the forms:

$$\begin{cases} v_{\pm}^i\left(\bar{x}, s\right) = \lim\limits_{\Delta s = 0\pm} \dfrac{1}{\Delta s} E\left\{ \left[ x^i\left(s + \Delta s\right) - x^i\left(s\right) \right] \mid \bar{x}\left(s\right) \right\} \\ w_{\pm}^{ij}\left(\bar{x}, s\right) = \lim\limits_{\Delta s = 0\pm} \dfrac{1}{\Delta s} E\left\{ \left[ x^i\left(s + \Delta s\right) - x^i\left(s\right) \right] \left[ x^j\left(s + \Delta s\right) - x^j\left(s\right) \right] \mid \bar{x}\left(s\right) \right\} \end{cases} \tag{1-8}$$

Let us remark that the quadratic form built with the components $w^{ij}$ is positive definite. This leads to the important conclusion that the determinants $\left| w^{ij} \right|$ must be different from zero and thus that the inverse matrix $\left(w_{ij}\right) = \left(w^{ij}\right)^{-1}$ exists.

We shall henceforth suppose that the probability density $p\left(\bar{x}, s\right)$, the conditional probability density $p\left(\bar{x}_1, s_1 \mid \bar{x}_0, s_0\right)$, the drift vectors $v_{\pm}^i\left(\bar{x}, s\right)$ and the diffusion tensors $w_{\pm}^{ij}\left(\bar{x}, s\right)$ possess all regularity properties necessary and sufficient for the further development of the theory. From the equation (1-5) and from the definitions (1-8) we can then deduce that the probability density $p\left(\bar{x}, s\right)$ verifies the Fokker-Planck equations:

$$\frac{\partial p\left(\bar{x}, s\right)}{\partial s} + \frac{\partial \left[ v_{\pm}^i\left(\bar{x}, s\right) p\left(\bar{x}, s\right) \right]}{\partial x^i} \mp \frac{\partial^2 \left[ w_{\pm}^{ij}\left(\bar{x}, s\right) p\left(\bar{x}, s\right) \right]}{\partial x^i\, \partial x^j} = 0 \tag{1-9}$$

and that the conditional probability density $p\left(\bar{x}_1, s_1 \mid \bar{x}_0, s_0\right)$ verifies the first Kolmogorov equations:

$$\frac{\partial p\left(\overline{x}_1,s_1 \mid \overline{x}_0,s_0\right)}{\partial s_0}+v_{\pm}^{i}(\overline{x}_0,s_0)\frac{\partial p\left(\overline{x}_1,s_1 \mid \overline{x}_0,s_0\right)}{\partial x_0^{i}}$$
$$\pm w_{\pm}^{ij}(x_0,s_0)\frac{\partial^2 p\left(\overline{x}_1,s_1 \mid \overline{x}_0,s_0\right)}{\partial x_0^{i}\,\partial x_0^{j}}=0 \tag{1-10}$$

and the second Kolmogorov equations:

$$\frac{\partial p\left(\overline{x}_1,s_1 \mid \overline{x}_0,s_0\right)}{\partial s_1}+\frac{\partial\left[v_{\pm}^{i}\left(\overline{x}_1,s_1\right)p\left(\overline{x}_1,s_1 \mid \overline{x}_0,s_0\right)\right]}{\partial x_1^{i}}$$
$$\mp\frac{\partial^2\left[w_{\pm}^{ij}\left(\overline{x}_1,s_1\right)p\left(\overline{x}_1,s_1 \mid \overline{x}_0,s_0\right)\right]}{\partial x_1^{i}\,\partial x_1^{j}}=0 \tag{1-11}$$

Let us remark that the Chapman-Kolmogorov equation (1-5) can be generalized to the case where either $s_1=s_0$ or $s_2=s_1$ provided that we impose the condition:

$$p\left(\overline{x}_1,s_0 \mid \overline{x}_0,s_0\right)=\delta\left(\overline{x}_1-\overline{x}_0\right) \tag{1-12}$$

where $\delta$ is the Dirac function.

So the first Kolmogorov equations are partial differential equations of parabolic type in the variables $\overline{x}_0$ and $s_0$ for the unknown function $p\left(\overline{x}_1,s_1 \mid \overline{x}_0,s_0\right)$ with the initial condition (1-12).

In the same way, the second Kolmogorov equations are partial differential equations of parabolic type in the variables $\overline{x}_1$ and $s_1$ for the unknown function $p\left(\overline{x}_1,s_1 \mid \overline{x}_0,s_0\right)$ with the same initial condition (1-12).

It can be proved that the two stochastic difference equations differing by the sign of $\Delta s$:

$$x^{i}\left(s+\Delta s\right)=x^{i}\left(s\right)+a_{\pm}^{i}\left(\overline{x}\left(s\right),s\right)\Delta s+b_{\pm}^{ij}\left(\overline{x}\left(s\right),s\right)\left[z_j^{\pm}\left(s+\Delta s\right)-z_j^{\pm}\left(s\right)\right] \tag{1-13}$$

where the vector stochastic processes $z_r^{\pm}$ are independent generalized Wiener processes that are submitted to the conditions:

$$\begin{cases} E\left\{ z_r^{\pm}\left(s_{k+1}\right)-z_r^{\pm}\left(s_k\right) \middle| \overline{x}\left(s_k\right)\right\}=0 \\ E\left\{\left[ z_r^{\pm}\left(s_{k+1}\right)-z_r^{\pm}\left(s_k\right)\right]^2 \middle| \overline{x}\left(s_k\right)\right\}=T_r^{\pm}\left(s_{k+1}\right)-T_r^{\pm}\left(s_k\right) \\ E\left\{\left[ z_r^{\pm}\left(s_{k+1}\right)-z_r^{\pm}\left(s_k\right)\right]\left[ z_s^{\pm}\left(s_{k+1}\right)-z_s^{\pm}\left(s_k\right)\right] \middle| \overline{x}\left(s_k\right)\right\}=0 \quad \text{if } r\neq s \end{cases} \tag{1-14}$$

determine diffusion processes whose drift vectors and diffusion tensors are given by:

$$\begin{cases} v_{\pm}^{i}\left(\overline{x}\left(s\right),s\right)=a_{\pm}^{i}\left(\overline{x}\left(s\right),s\right) \\ w_{\pm}^{ij}\left(\overline{x}\left(s\right),s\right)=\dfrac{1}{2}\displaystyle\sum_{r=0}^{3} b_{\pm}^{ir}\left(\overline{x}\left(s\right),s\right)b_{\pm}^{jr}\left(\overline{x}\left(s_0\right),s_0\right)\frac{d\,T_k^{\pm}\left(s\right)}{ds} \end{cases} \tag{1-15}$$

We shall henceforth restrict our attention to the diffusions processes so defined. Let us then mention that the difference equation (1-13) can be used to prove that although (almost) all sample functions $\bar{x}(s)$ are continuous they have (almost) nowhere a derivative with respect to the parameter s.

We define now the stochastic derivatives by the formulae:

$$
\begin{aligned}
D_{\pm}f &= \lim_{h=\pm 0} E\left\{ \frac{f\left(\bar{x}(s+h), s+h\right) - f\left(\bar{x}(s), s\right)}{h} \;\middle|\; x(s) \right\} \\
&= \frac{\partial f(\bar{x},s)}{\partial s} + v_{\pm}^{i}(\bar{x},s)\partial_i f(\bar{x},s) \pm w_{\pm}^{ij}(\bar{x},s)\partial_{ij}^{2} f(\bar{x},s)
\end{aligned}
\tag{1-16}
$$

with the particular case:

$$
D_{\pm}x^{i}(s) = v_{\pm}^{i}\left(x(s), s\right) \tag{1-17}
$$

The two stochastic derivatives can be combined to give:

$$
E\left\{ f(b)g(b) - f(a)g(a) \right\} = \int_{a}^{b} E\left\{ g(s)D_{+}f(s) + f(s)D_{-}g(s) \right\} ds \tag{1-18}
$$

that can be used to prove that the drift vectors $v_{\pm}^{i}$ and diffusion tensors $w_{\pm}^{ij}$ are not independent but obey the relations:

$$
\begin{cases}
w_{+}^{ij}(\bar{x},s) = w_{-}^{ij}(\bar{x},s) = w^{ij}(\bar{x},s) \\
v_{+}^{i}(\bar{x},s) - v_{-}^{i}(\bar{x},s) = \dfrac{2}{p(\bar{x},s)} \partial_j \left[ p(\bar{x},s) w^{ij}(\bar{x},s) \right]
\end{cases}
\tag{1-19}
$$

where we have denoted by $w^{ij}(\bar{x},s)$ the common value of $w_{+}^{ij}(\bar{x},s)$ and $w_{-}^{ij}(\bar{x},s)$. If we now define the auxiliary vectors $u^{i}(\bar{x},s)$ and $v^{i}(\bar{x},s)$ by the relations:

$$
\begin{cases}
u^{i}(\bar{x},s) = \left[ v_{+}^{i}(\bar{x},s) - v_{-}^{i}(\bar{x},s) \right] / 2 \\
v^{i}(\bar{x},s) = \left[ v_{+}^{i}(\bar{x},s) + v_{-}^{i}(\bar{x},s) \right] / 2
\end{cases}
\tag{1-20}
$$

we see that the relations (1-19) lead to the equation:

$$
p(\bar{x},s) u^{i}(\bar{x},s) = \partial_j \left[ p(\bar{x},s) w^{ij}(\bar{x},s) \right] \tag{1-21}
$$

and that the addition of the two forms of the Fokker-Planck equations (1-9) members to members leads to the continuity equation:

$$
\frac{\partial p(\bar{x},s)}{\partial s} + \partial_i \left[ p(\bar{x},s) v^{i}(\bar{x},s) \right] = 0 \tag{1-22}
$$

There is thus no difficulty for writing down the definitions and equations describing a diffusion process in a Minkowski space, but we did not yet care about the principles of relativity. By remembering the basic properties of covariant and contravariant vectors and tensors and by having a quick look at the proposed equations, and especially at the Fokker-Planck and Kolmogorov equations, we are easily convinced that their relativistic invariance will be a direct consequence of the fact that the drift vectors $v_{\pm}^{i}(\bar{x},s)$ are contravariant vectors and that the diffusion tensor $w^{ij}(\bar{x},s)$ is a two times contravariant tensor with respect to the generalized Lorentz transformations. Designating the new co-ordinates (and all functions expressed by means of them) by upper case letters ([1]), we know that those Lorentz transformations are defined by linear relations with constant coefficients $X_k^i$ as in:

$$X^i = X_k^i \, x^k \tag{1-23}$$

Introducing the new co-ordinates in the definitions (1-8) leads to:

$$\begin{aligned} V_\pm^r(\bar{X},s) &= \lim_{\Delta s=\pm 0} \frac{1}{\Delta s} E\left\{ \left[ X^r(s+\Delta s) - X^r(s) \right] \middle| \bar{X}(s) \right\} \\ &= X_i^r \lim_{\Delta s=\pm 0} \frac{1}{\Delta s} E\left\{ \left[ x^i(s+\Delta s) - x^i(s) \right] \middle| \bar{x}(s) \right\} \\ &= X_i^r \, v_\pm^i(\bar{x},s) \end{aligned} \tag{1-24}$$

showing that the drift vectors $v_\pm^i(\bar{x},s)$ are contravariant vectors. Note that the points $\bar{x}(s)$ and $\bar{X}(s)$ coincide and that their different writings simply express the fact that they are referenced by different systems of co-ordinates. This explains why the conditional expectations with respect to $\bar{x}(s)$ and $\bar{X}(s)$ are equal. Similarly, we obtain from the definitions (1-8):

$$W^{rs}(\bar{X},s) = X_i^r X_j^s \, w^{ij}(\bar{x},s) \tag{1-25}$$

showing that the diffusion tensor $w^{ij}(\bar{x},s)$ is a two times contravariant tensor.

There is also an important relativistic condition involving the dynamical parameter s. This parameter seems to be unique in non-relativistic mechanics because the time t is always chosen for playing this role. This is not so in relativistic mechanics where it is even possible to use different parameters for different particles. So we must also prove that all equations describing a diffusion process in a Minkowski space are in some sense independent on the chosen parameter. We shall therefore consider a new parameter S that is a continuous function of the old parameter s and whose derivative is everywhere positive and finite. We shall designate hereafter by uppercase letters the co-ordinates and all functions considered as depending on the parameter S. So we shall write:

$$X^i(S) = x^i(s) \tag{1-26}$$

The definitions (1-8) lead to:

[1] The same symbols written with uppercase letters like U, V, W, ... will be used repeatedly to designate the new values of the corresponding symbols written with lowercase letters like u, v, w, ...either after a Lorentz transformation or after a change of dynamical parameter or after a change of variables. We hope that the context will be sufficiently clear to avoid possible confusion, but imagining three completely different sets of temporary notations is cumbersome and can do after all more harm than good.

$$
\begin{aligned}
V_{\pm}^{i}(\overline{x},S)dS &= \lim_{\Delta S=\pm 0} E\left\{ X^{i}(S+\Delta S)-X^{i}(S)\middle|\overline{X}(S)\right\} \\
&= \lim_{\Delta s=\pm 0} E\left\{ x^{i}(s+\Delta s)-x^{i}(s)\middle|\overline{x}(s)\right\} \\
&= v_{\pm}^{i}(\overline{x},s)ds
\end{aligned}
\tag{1-27}
$$

and to:

$$
W^{ij}(\overline{x},S)dS = w^{ij}(\overline{x},s)ds \tag{1-28}
$$

We shall present later convincing arguments indicating that the theory of the relativistic diffusion process is somewhat similar to the theory of the non-relativistic stationary diffusion process. This one is characterized by the fact that the drift vectors $v_{\pm}^{i}$ (and the diffusion tensor $w^{ij}$) are independent on the time t, with as consequences that the probability density $p(\overline{x},t)$ is also actually independent on the time t and that the time variable t is separable from the space variables x, y and z in the Fokker-Planck and Kolmogorov equations. In the context of a relativistic diffusion process, it is not possible to maintain that the drift vectors $v_{\pm}^{i}$ and the diffusion tensor $w^{ij}$ can be independent on the dynamical parameter s because of the relations (1-27) and (1-28), but it is however possible to transpose many interesting properties from the non-relativistic into the relativistic domain if we suppose that the drift vectors $v_{\pm}^{i}(\overline{x},s)$ and $V_{\pm}^{i}(\overline{x},S)$ and the diffusion tensors $w^{ij}(\overline{x},s)$ and $W^{ij}(\overline{x},S)$ have the forms:

$$
\begin{cases}
v_{\pm}^{i}(\overline{x},s) = v_{\pm}^{i}(\overline{x})\dot{g}(s) \\
V_{\pm}^{i}(\overline{x},S) = V_{\pm}^{i}(\overline{x})\dot{G}(S)
\end{cases}
\tag{1-29}
$$

and:

$$
\begin{cases}
w^{ij}(\overline{x},s) = w^{ij}(\overline{x})\dot{g}(s) \\
W^{ij}(\overline{x},S) = W^{ij}(\overline{x})\dot{G}(S)
\end{cases}
\tag{1-30}
$$

where $g(s)$ and $G(S)$ designate strictly increasing functions of their arguments s or S that correspond to each other in the direct transform $S=S(s)$ or in its inverse transform $s=s(S)$, where the dot indicates a derivative with respect to s or S as the case may be and where we have used lowercase or uppercase letters to obey the naming convention for the drift vectors $v_{\pm}^{i}(\overline{x})$ and $V_{\pm}^{i}(\overline{x})$ and for the diffusion tensors $w^{ij}(\overline{x})$ and $W^{ij}(\overline{x})$ in spite of the fact that they are actually identical even if they have been defined in the contexts of the two different dynamical parameters s and S.

If we now replace the drift vectors $v_{\pm}^{i}(\overline{x},s)$ and the diffusion tensor $w^{ij}(\overline{x},s)$ by the expressions proposed in (1-29) and (1-30) we can rewrite the Fokker-Planck and Kolmogorov equations in the forms:

$$
\frac{1}{\dot{g}(s)}\frac{\partial p(\overline{x},s)}{\partial s} + \frac{\partial}{\partial x^{i}}\left[ p(\overline{x},s)v_{\pm}^{i}(\overline{x})\right] \mp \frac{\partial^{2}}{\partial x^{i}\partial x^{j}}\left[ p(\overline{x},s)w^{ij}(\overline{x})\right] = 0 \tag{1-31}
$$

and

$$\frac{1}{\dot{g}(s_0)}\frac{\partial p(\overline{x}_1,s_1\,|\,\overline{x}_0,s_0)}{\partial s_0}+v_{\pm}^{i}(\overline{x}_0)\frac{\partial p(\overline{x}_1,s_1\,|\,\overline{x}_0,s_0)}{\partial x_0^i}$$
$$\pm w^{ij}(\overline{x}_0)\frac{\partial^2 p(\overline{x}_1,s_1\,|\,\overline{x}_0,s_0)}{\partial x_0^i\,\partial x_0^j}=0 \quad (1\text{-}32)$$

$$\frac{1}{\dot{g}(s_1)}\frac{\partial p(\overline{x}_1,s_1\,|\,\overline{x}_0,s_0)}{\partial s_1}+\frac{\partial}{\partial x^i}\Big[\,p(\overline{x}_1,s_1\,|\,\overline{x}_0,s_0)\,v_{\pm}^{i}(\overline{x})\Big]$$
$$\mp\frac{\partial^2}{\partial x^i\,\partial x^j}\Big[\,p(\overline{x}_1,s_1\,|\,\overline{x}_0,s_0)\,w^{ij}(\overline{x})\Big]=0 \quad (1\text{-}33)$$

If we define in a natural manner the auxiliary vectors:

$$\begin{cases} u^i(\overline{x})=\Big[v_{+}^{i}(\overline{x})-v_{-}^{i}(\overline{x})\Big]\Big/2 \\ v^i(\overline{x})=\Big[v_{+}^{i}(\overline{x})+v_{-}^{i}(\overline{x})\Big]\Big/2 \end{cases} \quad (1\text{-}34)$$

we can obviously write

$$\begin{cases} u^i(\overline{x},s)=u^i(\overline{x})\,\dot{g}(s) \\ v^i(\overline{x},s)=v^i(\overline{x})\,\dot{g}(s) \end{cases} \quad (1\text{-}35)$$

and it is clear that the equations (1-21) and (1-22) can be written in the forms:

$$p(\overline{x},s)\,u^i(\overline{x})=\partial_j\Big[p(\overline{x},s)\,w^{ij}(\overline{x})\Big] \quad (1\text{-}36)$$

and:

$$\frac{1}{\dot{g}(s)}\frac{\partial p(\overline{x},s)}{\partial s}+\partial_i\Big[p(\overline{x},s)\,v^i(\overline{x})\Big]=0 \quad (1\text{-}37)$$

So everything is fine from the mathematical point of view. We must however remark that quantum theory is actually a physical theory. We can object for example that if $ct$ is considered as a random variable in relativistic quantum theory, it is also such a variable in non-relativistic quantum theory that we have previously developed by supposing on the contrary that t could be used as the dynamical parameter. We can write however:

$$E\Big\{\Big[\,t(s)-t(s_0)\Big]^2\,\Big|\,\overline{x}(s_0)\Big\}=\frac{1}{c^2}E\Big\{\Big[\,ct(s)-ct(s_0)\Big]^2\,\Big|\,\overline{x}(s_0)\Big\}$$
$$\approx\frac{2}{c^2}\,w^{00}\big(\overline{x}(s_0),s_0\big)\,|\,s-s_0| \quad (1\text{-}38)$$

showing that the conditional expectation in the left-hand side member converges to 0 when c increases indefinitely to $\infty$. At this limit, the variable t is no longer a random variable and can be used safely as the dynamical parameter. We can object conversely that if t is not a random variable in non-relativistic theory, then ct is perhaps not such a variable in relativistic quantum theory either. But according to the Lorentz transformation (1-23), we must write:

$$X^0=X_0^0x^0+X_1^0x^1+X_2^0x^2+X_3^0x^3 \quad (1\text{-}39)$$

and claiming that the variables $x^1$, $x^2$ and $x^3$ are random variables while this is not the case for the variables $x^0$ and $X^0$ is a mere contradiction.

You may still claim that our definition of a diffusion process in a Minkowski space is not yet a definition of a fully relativistic diffusion process. You may present as an argument for this thesis that we proposed no conditions to forbid the presence in the definitions (1-8) of terms for which the points $\overline{x}_0$ and $\overline{x}_1$ lie outside each other light cones and may not be points on a normal trajectory. But classical theory already allows us to join such two points by a broken curve with partial arcs along which relativity is nowhere violated. The price to pay is to concede that time t must be a decreasing function of the dynamical parameter s along some special arcs while it is normally an increasing function of s. There is no objection against this picture of the trajectory of a particle because the special arcs can be considered as normal arcs for the corresponding antiparticle and the turning points can then be interpreted as the creation or annihilation of a pair. The very irregular nature of the trajectories as sample functions of a diffusion process can only be in favour of this point of view. We do not intend to continue this discussion further and we prefer to proceed with the development of the relativistic quantum theory, hoping that its results will support it.

## 2. Nelson first equation

We now intend to find the relativistic versions of the Nelson equations that will be used to determine the vectors $u^i\left(\overline{x},s\right)$ and $v^i\left(\overline{x},s\right)$. The first Nelson equation has a purely probabilistic origin. We shall see that it is possible to transpose many interesting results from the theory of the non-relativistic stationary diffusion processes to the theory of the relativistic diffusion processes if we suppose that the drift vectors $v^i_{\pm}\left(\overline{x},s\right)$ and the diffusion tensor $w^{ij}\left(\overline{x},s\right)$ are given by (1-29) and (1-30). A direct consequence is that the probability density $p\left(\overline{x},s\right)$ is actually independent on the dynamical parameter s. Then the conservation equation (1-22) loses its first term and if we define the matrix $\left(w_{ij}\right)$ as being the inverse of the matrix $\left(w^{ij}\right)$, we can then write:

$$\begin{aligned} 0 &= \partial_i\left[\,p\left(\overline{x}\right)v^i\left(\overline{x}\right)\right] \\ &= \partial_i\left[\,p\left(\overline{x}\right)w^{ij}\left(\overline{x}\right)w_{jk}\left(\overline{x}\right)v^k\left(\overline{x}\right)\right] \\ &= w_{jk}\left(\overline{x}\right)v^k\left(\overline{x}\right)\partial_i\left[\,p\left(\overline{x}\right)w^{ij}\left(\overline{x}\right)\,\right]+p\left(\overline{x}\right)w^{ij}\left(\overline{x}\right)\partial_i\left[\,w_{jk}\left(\overline{x}\right)v^k\left(\overline{x}\right)\,\right] \\ &= p\left(\overline{x}\right)\left\{w_{jk}\left(\overline{x}\right)u^j\left(\overline{x}\right)v^k\left(\overline{x}\right)+w^{ij}\left(\overline{x}\right)\partial_i\left[w_{jk}\left(\overline{x}\right)v^k\left(\overline{x}\right)\right]\right\} \end{aligned} \qquad (2\text{-}1)$$

and we so arrive at the so-called first Nelson equation:

$$w_{ij}\left(\overline{x}\right)u^i\left(\overline{x}\right)v^j\left(\overline{x}\right)+w^{ij}\left(\overline{x}\right)\partial_i\left[w_{jk}\left(\overline{x}\right)v^k\left(\overline{x}\right)\right]=0 \qquad (2\text{-}2)$$

## 3. Nelson second equations

The Nelson second equations are deduced from the stochastic variational problem applied to the Lagrangian:

$$L(\overline{x},\overline{v}_+,\overline{v}_+)=mc\sqrt{g_{ij}\left[v^i\left(\overline{x},s\right)v^j\left(\overline{x},s\right)+u^i\left(\overline{x},s\right)u^j\left(\overline{x},s\right)\right]}-qA_j\left(\overline{x}\right)v^j\left(\overline{x},s\right) \qquad (3\text{-}1)$$

that is at the same time a relativistic version of the one used in the non-relativistic quantum theory and a stochastic version of the one used in the classical relativity theory. Here above m is the mass of the particle and q is its charge. The covariant vector $A_j(\bar{x})$ is the electromagnetic potential defined in a Minkowski space by its components:

$$\begin{cases} A^0(\bar{x}) = A_0(\bar{x}) = V(\bar{x})/c \\ A^1(\bar{x}) = -A_1(\bar{x}) = A_x(\bar{x}) \\ A^2(\bar{x}) = -A_2(\bar{x}) = A_y(\bar{x}) \\ A^3(\bar{x}) = -A_3(\bar{x}) = A_z(\bar{x}) \end{cases} \tag{3-2}$$

where $V(\bar{x})$ is the scalar electric potential and where $\bar{A}(\bar{x})$ is the vector magnetic potential. It is important to note that the electric and magnetic fields and potentials are normally functions that depend on the four co-ordinates of the point $\bar{x}$ in a Minkowski space, but that surely do not depend on the dynamical parameter s. Supposing the contrary would amount to admit that the values of those quantities at a point $\bar{x}$ are not unequivocally determined but can still vary as a function of the extra parameter s, what is physically unsound

Before proceeding we shall first define the so-called normalized vectors $\tilde{u}^i(\bar{x})$ and $\tilde{v}^i(\bar{x})$. If we use the formulae:

$$\begin{cases} |\, v(x)\,|^2 = g_{ij}\left[ v^i(\bar{x}) v^j(\bar{x}) + u^i(\bar{x}) u^j(\bar{x}) \right] \\ |\, v(x,s)\,|^2 = g_{ij}\left[ v^i(\bar{x},s) v^j(\bar{x},s) + u^i(\bar{x},s) u^j(\bar{x},s) \right] \end{cases} \tag{3-3}$$

to evaluate the scalars $|\, v(\bar{x})\,|$ and $|\, v(\bar{x},s)\,|$ we obviously have:

$$|\, v(\bar{x},s)\,| = |\, v(\bar{x})\,|\, \dot{g}(s) \tag{3-4}$$

so that the vectors $\tilde{u}^i(\bar{x})$ and $\tilde{v}^i(\bar{x})$ defined by the equivalent relations:

$$\begin{cases} \tilde{u}^i(\bar{x}) = c\, u^i(\bar{x},s) / |\, v(\bar{x},s)\,| = c\, u^i(\bar{x}) / |\, v(\bar{x})\,| \\ \tilde{v}^i(\bar{x}) = c\, v^i(\bar{x},s) / |\, v(\bar{x},s)\,| = c\, v^i(\bar{x}) / |\, v(\bar{x})\,| \end{cases} \tag{3-5}$$

are obviously independent on the dynamic parameter. The chosen symbol $|\, v\,|$ helps to remember that it will always designate a non-negative quantity and the factor c has been introduced for the only reason that it makes identical the dimensions of the corresponding original and normalized variables. A simple calculation gives immediately:

$$\begin{aligned} |\, \tilde{v}(\bar{x})\,|^2 &= g_{ij}\left[ \tilde{v}^i(\bar{x}) \tilde{v}^j(\bar{x}) + \tilde{u}^i(\bar{x}) \tilde{u}^j(\bar{x}) \right] \\ &= c^2 \end{aligned} \tag{3-6}$$

Using the rules that allow us to raise or to lower freely indices of a covariant or contravariant vector we can write:

$$\frac{\partial L}{\partial u^k} = m c \frac{g_{ik}\, u^i\left(\bar{x},s\right)}{\left|\, v\left(\bar{x},s\right)\right|} \tag{3-7}$$

$$= m\, \tilde{u}_k\left(\bar{x}\right)$$

and:

$$\frac{\partial L}{\partial v^k} = m\, \tilde{v}_k\left(\bar{x}\right) - q\, A_k\left(\bar{x}\right) \tag{3-8}$$

So the Euler-Lagrange equations presented in their form:

$$\frac{\partial}{\partial t}\frac{\partial L}{\partial v^k} + \sum_{i=1}^{m} v^i \frac{\partial}{\partial x^i}\frac{\partial L}{\partial v^k} - \sum_{i=1}^{m} u^i \frac{\partial}{\partial x^i}\frac{\partial L}{\partial u^k} - \sum_{i=1}^{m}\sum_{j=1}^{m} w^{ij}\frac{\partial^2}{\partial x^i \partial x^j}\frac{\partial L}{\partial u^k} = \frac{\partial L}{\partial x^k} \tag{3-9}$$

lead to:

$$\begin{aligned}&\frac{\partial}{\partial s}\left[m\, \tilde{v}_k\left(\bar{x}\right) - q\, A_k\left(\bar{x}\right)\right] + v^i\left(\bar{x},s\right)\partial_i\left[m\, \tilde{v}_k\left(\bar{x}\right) - q\, A_k\left(\bar{x}\right)\right]\\ &\quad - u^i\left(\bar{x},s\right)\partial_i\left[m\, \tilde{u}_k\left(\bar{x}\right)\right] - w^{ij}\left(\bar{x},s\right)\partial_{ij}\left[m\, \tilde{u}_k\left(\bar{x}\right)\right] = -q\, v^i\left(\bar{x},s\right)\partial_k A_i\left(\bar{x}\right)\end{aligned} \tag{3-10}$$

Remember that the normalized variables $\tilde{u}_i\left(\bar{x}\right)$ and $\tilde{v}_i\left(\bar{x}\right)$ that arose automatically in the relations (3-7) and (3-8)) are independent on the dynamical parameter s, so that the Euler-Lagrange equations that we shall henceforth call the second Nelson equations become:

$$\begin{aligned}&m\left[\, v^i\left(\bar{x},s\right)\partial_i \tilde{v}_k\left(\bar{x}\right) - u^i\left(\bar{x},s\right)\partial_i \tilde{u}_k\left(\bar{x}\right) - w^{ij}\left(\bar{x},s\right)\partial_{ij}^2 \tilde{u}_k\left(\bar{x}\right)\right]\\ &\qquad = q\, v^i\left(\bar{x},s\right)\left[\partial_i\, A_k\left(\bar{x}\right) - \partial_k\, A_i\left(\bar{x}\right)\right]\end{aligned} \tag{3-11}$$

We finally note that the expression:

$$\frac{\partial L}{\partial \tilde{v}^i} = m\, \tilde{v}_i\left(\bar{x}\right) - q\, A_i\left(\bar{x}\right) \tag{3-12}$$

must be a gradient as it results from the theory of the stochastic variational principle.

## 4. Diffusion tensor

In non-relativistic quantum theory, we were not able to present a principle from which it was possible to deduce the expression:

$$w^{ij} = \frac{h}{4\pi m} g^{ij} \tag{4-1}$$

for the diffusion tensor $w^{ij}$ and this form was justified only by comparing the Nelson equations with relations deduced from the Schrödinger equation. We are facing the same problem in relativistic quantum theory and the first idea that comes to mind to determine the form of this tensor is to apply the same procedure to the Klein-Gordon equation. This leads to:

$$w^{ij} = -\frac{h}{4\pi m} g^{ij} \tag{4-2}$$

Unfortunately, this very simple form for the tensor $w^{ij}$ is not acceptable because $w^{00}$ must be non-negative according to its definition. So we must find a correspondence other than (4-2). After many hesitations, we have decided to propose the definition:

$$w^{ij}(\bar{x}) = \frac{h}{4\pi m}\left[2\frac{v(\bar{x})^i v(\bar{x})^j}{|v(\bar{x})|^2} - g^{ij}\right] \tag{4-3}$$

for the reason that the diffusion tensor $w^{ij}(\bar{x})$ so defined possesses useful properties. First it can be proved that the quadratic form $w^{ij}(\bar{x})a_i a_j$ is positive definite as it is imposed to it.

Then the inverse $w_{ij}(\bar{x})$ of the diffusion tensor $w^{ij}(\bar{x})$ is simply given by:

$$w_{ij}(\bar{x}) = \frac{4\pi m}{h}\left[2\frac{v_i(\bar{x})v_j(\bar{x})}{|v(\bar{x})|^2} - g_{ij}\right] \tag{4-4}$$

as this results from a simple calculation leading to the identity:

$$w_{ik}(\bar{x})w^{jk}(\bar{x}) = \delta_i^j \tag{4-5}$$

Finally, we also have:

$$w_{ij}(\bar{x})v^j(\bar{x}) = \frac{4\pi m}{h}v_i(\bar{x}) \tag{4-6}$$

and this allows writing the first Nelson equation (2-2) in the simpler form:

$$u^j(\bar{x})v_j(\bar{x}) + w^{ij}(\bar{x})\partial_i v_j(\bar{x}) = 0 \tag{4-7}$$

A last good point is that we have at the non-relativistic limit:

$$\lim_{c=\infty}\tilde{w}^{ij} = -\frac{h}{4\pi m}g^{ij} \quad \text{for } i \neq 0 \neq i \tag{4-8}$$

We feel obliged to say that there exists unfortunately a bad point against the definitions (4-3). It is expressed by the inequality:

$$\begin{aligned} E\left\{g_{ij}\left(x_1^i - x_0^i\right)\left(x_1^j - x_0^j\right) \middle| \bar{x}_0\right\} &= g_{ij}E\left\{\left(x_1^i - x_0^i\right)\left(x_1^j - x_0^j\right) \middle| \bar{x}_0\right\} \\ &= \frac{h}{4\pi m}g_{ij}\left[2\frac{v(\bar{x})^i v(\bar{x})^j}{|v(\bar{x})|^2} - g^{ij}\right]|\Delta s| \\ &= -\frac{h}{2\pi m}|\Delta s| \\ &< 0 \end{aligned} \tag{4-9}$$

Remember that we have decided from the start to not eliminate the space-like displacements $x_1^i - x_0^i$ from the conditional mathematical expectation, but the definition (4-3) of the diffusion tensor favours these with respect to the time-like displacements. The problem is more difficult than expected, because another choice of the diffusion tensor that would lead to the supposedly better condition $g_{ij}w^{ij} > 0$ does not help. We know indeed that its statistical interpretation does not rule out space-like displacements.

Everything looks fine at first sight because we can solve the Nelson equations (2-2) and (3-11) with the definition (4-3) to determine first the drift vectors $u^i(\bar{x})$ and $v^i(\bar{x})$ or $v^i_{\pm}(\bar{x})$ in order then to write down and solve the Fokker-Planck and Kolmogorov equations. At closer inspection this appears to be a task more difficult than in the non-relativistic case because the second Nelson equation involves the drift vectors in their original form, but also in their normalized versions and in the unpleasant irrational denominator $|v(\bar{x})|$, but this amounts after all to a purely mathematical problem only. Anyway those equations are horribly intricate and are certainly not the kind we like if we are asked to solve them, if we can at all. We shall often cheat by guessing what the solutions look like instead of determining exactly what they are.

We think that it is high time to produce exhibits in illustration and defence of the relativistic stochastic quantum theory based on the finding that they will not lead to unsound results when it will be applied to the study of (an) especially simple problem(s). We cannot do this however without crossing the Rubicon and taking the momentous decision to follow the road paved with the simplifying hypotheses (1-29), (1-30) and (4-3). We hope that everything will run like clockwork.

## 5. Free particle

Before proceeding to the study of the very simple case of the free particle, we think that it is a good idea to explain how a problem stated in the relativistic context of the quantum mechanics can be solved by the methods of the relativistic stochastic theory.

We normally start from the specification of an electromagnetic potential $A_i(\bar{x})$ acting on a particle of mass m and charge q. In a first step, we must determine the vectors $u^i(\bar{x})$ and $v^i(\bar{x})$ that verify the normalization condition:

$$v_k(\bar{x})v^k(\bar{x}) + u_k(\bar{x})u^k(\bar{x}) = c^2 \tag{5-1}$$

and that with the value:

$$w^{ij}(\bar{x}) = \frac{h}{4\pi m}\left[2\frac{v^i(\bar{x})v^j(\bar{x})}{v_k(\bar{x})v^k(\bar{x})} - g^{ij}\right] \tag{5-2}$$

of the diffusion tensor are solutions of the two Nelson equations:

$$v_i(\bar{x})u^i(\bar{x}) + w^{ij}(\bar{x})\partial_i v_j(\bar{x}) = 0 \tag{5-3}$$

and:

$$\begin{aligned} m\left[v^i(\bar{x})\partial_i v_k(\bar{x}) - u^i(\bar{x})\partial_i u_k(\bar{x}) - w^{ij}(\bar{x})\partial^2_{ij}u_i(\bar{x})\right] \\ = q\,v^i(\bar{x})\left[\partial_k A_i(\bar{x}) - \partial_i A_k(\bar{x})\right] \end{aligned} \tag{5-4}$$

At this point, we must call the attention to the fact that not all mathematical solutions of those equations are acceptable because the variational principle imposes the condition:

$$m\, v_i\left(\overline{x}\right) + q\, A_i\left(\overline{x}\right) = \partial_i G\left(\overline{x}\right) \tag{5-5}$$

for some scalar function $G\left(\overline{x}\right)$.

In a second step, we must solve the two Kolmogorov equations:

$$\begin{aligned}\frac{1}{\dot{g}\left(s_0\right)}\frac{\partial p\left(\overline{x},s\,|\,\overline{x}_0,s_0\right)}{\partial s_0} + v_{\pm}^{i}\left(\overline{x}_0\right)\frac{\partial p\left(\overline{x},s\,|\,\overline{x}_0,s_0\right)}{\partial x_0^{i}} \\ \pm w^{ij}\left(\overline{x}_0\right)\frac{\partial^2 p\left(\overline{x},s\,|\,\overline{x}_0,s_0\right)}{\partial x_0^{i}\,\partial x_0^{j}} = 0\end{aligned} \tag{5-6}$$

and:

$$\begin{aligned}\frac{1}{\dot{g}\left(s\right)}\frac{\partial p\left(\overline{x},s\,|\,\overline{x}_0,s_0\right)}{\partial s} + \frac{\partial}{\partial x^{i}}\left[\, p\left(\overline{x},s\,|\,\overline{x}_0,s_0\right) v_{\pm}^{i}\left(\overline{x}\right)\right] \\ \mp \frac{\partial^2}{\partial x^{i}\,\partial x^{j}}\left[\, p\left(\overline{x},s\,|\,\overline{x}_0,s_0\right) w^{ij}\left(\overline{x}\right)\right] = 0\end{aligned} \tag{5-7}$$

that together with the initial condition:

$$p\left(\overline{x},s_0 \,|\, \overline{x}_0,s_0\right) = \delta\left(\overline{x} - \overline{x}_0\right) \tag{5-8}$$

determine the conditional probability density $p\left(\overline{x},s\,|\,\overline{x}_0,s_0\right)$. If necessary, we must also solve the Fokker-Planck equation:

$$\frac{\partial}{\partial x^{i}}\left[\, p\left(\overline{x}\right) v_{\pm}^{i}\left(\overline{x}\right)\right] \mp \frac{\partial^2}{\partial x^{i}\,\partial x^{j}}\left[\, p\left(\overline{x}\right) w^{ij}\left(\overline{x}\right)\right] = 0 \tag{5-9}$$

that determines the probability density $p\left(\overline{x}\right)$. We must finally examine the pertinent mathematical properties of those normal and conditional probability densities and suggest for them possible physical interpretations that will hopefully justify the claim that the conditional probability density $p\left(\overline{x},s\,|\,\overline{x}_0,s_0\right)$ is a safely recommended substitute to the orthodox wave function.

We now intend to solve the problem of the free particle characterized by the fact that the force represented by the right-hand side member of equation (5-4) is identically zero. It is easy to see that a possible solution of the Nelson equations is given by:

$$u^{i} = 0 \tag{5-10}$$

and:

$$\begin{cases} v^{0} = c \\ v^{i} = 0 \quad \text{for } i \neq 0 \end{cases} \tag{5-11}$$

Formula (5-2) leads indeed to:

$$w^{ij} = \frac{h}{4\pi m}\begin{cases} 0 & \text{if } i \neq i \\ 1 & \text{if } i{=}j \end{cases} \tag{5-12}$$

and it is then a trivial task to verify that these values satisfy the normalization condition (5-1), the Nelson equations (5-3) and (5-4) and that the condition (5-5) imposed to the vector $v_i(\overline{x})$ is verified if we choose $G(\overline{x}) = m\,c\,t$

A concern is about the behaviour of the solution with respect to Lorentz transformations. Of course, the theory has been developed with great care about relativistic invariance, but because of the special values chosen for the diffusion tensor $w^{ij}$ we can still wonder if everything is perfect and if the renunciation to the d' Alembertian operator does not lead to physically unsound results.

We shall say that the free particle in at rest in the system f the co-ordinates $x^i$ where the vectors $u^i$ and $v^i$ and diffusion tensor $w^{ij}$ are given respectively by (5-10), (5-11) and (5-12). This is the simplest case, but we shall pay attention to the more general systems of co-ordinates $X^i$ where the free particle is moving with a constant translation velocity. The new co-ordinates are obviously related to the old co-ordinates by a Lorentz transformation that has the form:

$$X^i = X^i_j x^j \tag{5-13}$$

We know that the coefficients $X^i_j$ are constants that depend only on the translation velocity, but we are otherwise not interested by their exact definitions.

We shall designate by $U^i$, $V^i$ $W^{ij}$ the corresponding vectors and diffusion tensor in this new system and we shall also designate by $P(X, s \mid X_0, s_0)$ the conditional probability density related to it. Using the transforms:

$$\begin{cases} U^i = X^i_k u^k \\ V^i = X^i_k v^k \\ W^{ij} = X^i_k X^j_h w^{kh} \end{cases} \tag{5-14}$$

we easily find that all the components of the vector $U^i$ are equal to zero and that the components of the vector $V^i$ and of the diffusion tensor $W^{ij}$ are constants. The corresponding conditional probability density $p(\overline{X}, s \mid \overline{X}_0, s_0)$ is then the solution of the Kolmogorov equations:

$$\frac{1}{\dot{g}(s_0)} \frac{\partial P(\overline{X}, s \mid \overline{X}_0, s_0)}{\partial s_0} + V^i \frac{\partial P(\overline{X}, s \mid \overline{X}_0, s_0)}{\partial X^i_0} + W^{ij} \frac{\partial^2 P(\overline{X}, s \mid \overline{X}_0, s_0)}{\partial X^i_0 \partial X^j_0} = 0 \tag{5-15}$$

$$\frac{1}{\dot{g}(s)} \frac{\partial P(\overline{X}, s \mid \overline{X}_0, s_0)}{\partial s} + \frac{\partial \left(P(\overline{X}, s \mid \overline{X}_0, s_0) V^i\right)}{\partial X^i} - \frac{\partial^2 \left(P(\overline{X}, s \mid \overline{X}_0, s_0) W^{ij}\right)}{\partial X^i \partial X^j} = 0 \tag{5-16}$$

that verifies the initial condition:

$$P(\overline{X}, s_0 \mid \overline{X}_0, s_0) = \delta(\overline{X} - \overline{X}_0) \tag{5-17}$$

Let us mention that we have not forgotten the indices + or -, but that we have decided to omit them because we have $U^i = 0$ and because it simplifies somewhat the writing. Let us introduce the new variables:

$$\begin{cases} S = g - g_0 \\ Y^i = X^i - X_0^i - V^i\left(g - g_0\right) \end{cases} \tag{5-18}$$

where we have used the abbreviations:

$$\begin{cases} g = g\left(s\right) \\ g_0 = g\left(s_0\right) \end{cases} \tag{5-19}$$

We can see that the solution of the above pair of Kolmogorov equations together with their initial condition will be given by:

$$P\left(\bar{X},s \middle| \bar{X}_0,s_0\right) = Q\left(\bar{Y},S\right) \tag{5-20}$$

if and only if the function $Q\left(\bar{Y},S\right)$ is solution of the equation:

$$\frac{\partial Q\left(\bar{Y},S\right)}{\partial S} - W^{ij}\frac{\partial^2 Q\left(\bar{Y},S\right)}{\partial Y^i \partial Y^j} = 0 \tag{5-21}$$

and verifies the initial condition:

$$Q\left(\bar{Y},0\right) = \delta\left(\bar{Y}\right) \tag{5-22}$$

We can easily verify that the solution $Q\left(\bar{Y},S\right)$is:

$$Q\left(\bar{Y},S\right) = \sqrt{\frac{\det\left|W_{ij}\right|}{\left(4\pi S\right)^4}}\exp\left(-\frac{W_{ij}Y^iY^j}{4S}\right) \tag{5-23}$$

and that the solution $P\left(\bar{X},s \middle| \bar{X}_0,s_0\right)$ is then:

$$\begin{aligned} P\left(\bar{X},s \middle| \bar{X}_0,s_0\right) = &\sqrt{\frac{\det\left|W_{ij}\right|}{\left[4\pi\left(g-g_0\right)\right]^4}} \\ &\exp\left\{-\frac{W_{ij}\left[X^i - X_0^i - V^i\left(g-g_0\right)\right]\left[X^i - X_0^i - V^i\left(g-g_0\right)\right]}{4\left(g-g_0\right)}\right\} \end{aligned} \tag{5-24}$$

by taking into account the definition (5-20)), the changes of variables (5-18) and the solution (5-23) itself. This function is thus a solution of Kolmogorov equations (5-15) and (5-16) that verifies the initial condition (5-17) and is thus endowed with all the properties required for a conditional probability density.

In the system where the particle is at rest this conditional probability density takes the form:

$$p\left(\overline{x},s\,|\,\overline{x}_0,s_0\right)=\sqrt{\frac{m}{h\left(g-g_0\right)}}\exp\left\{-\frac{\pi m}{h}\frac{\left[x^0-x_0^0-c\left(g-g_0\right)\right]^2}{g-g_0}\right\}$$

$$\left(\sqrt{\frac{m}{h\left(g-g_0\right)}}\right)^3\exp\left[-\frac{\pi m}{h}\frac{\left(x^1-x_0^1\right)^2+\left(x^2-x_0^2\right)^2+\left(x^3-x_0^3\right)^2}{g-g_0}\right] \tag{5-25}$$

The solutions to the proposed Kolmogorov equations for the free particle possess several sympathetic properties, except perhaps that they are not zero for pairs of points $\overline{x}$ and $\overline{x}_0$ that lie outside each other light cone. We dare to propose an answer to this objection by invoking the possible creation or annihilation of pairs of particles and antiparticles. You would certainly applaud if you considered this explanation as a proof of the existence of positrons. We do not insist on this point but before leaving this subject, we find interesting to present a comparison of the values proposed by the solution (5-25) for two displacements equal to $c\tau$ in the t and x directions respectively for the same increment of the dynamical parameter $g\left(s\right)$. For the sake of simplicity, let us consider the three points $\overline{x}_0$, $\overline{x}_1$ and $\overline{x}_2$ with co-ordinates:

$$x_0^0=x_0^1=x_0^2=x_0^3=0 \tag{5-26}$$

$$\begin{cases}x_1^0=c\,\tau\\ x_1^1=x_1^2=x_1^3=0\end{cases} \tag{5-27}$$

$$\begin{cases}x_2^1=c\tau\\ x_2^0=x_2^2=x_2^3=0\end{cases} \tag{5-28}$$

and with also:

$$\begin{cases}g\left(s_0\right)=0\\ g\left(s_1\right)=g\left(s_2\right)=g\end{cases} \tag{5-29}$$

Let us remark that the interval $\left(\overline{x}_0,\overline{x}_1\right)$ is timelike, meaning that the straight line joining the two points is an acceptable relativistic trajectory, while the interval $\left(\overline{x}_0,\overline{x}_2\right)$ is spacelike, meaning that the straight line joining the two points is not an acceptable relativistic trajectory. From (5-25), we obtain:

$$p\left(\overline{x}_1,s\,|\,\overline{x}_0,s_0\right)=\left(\frac{m}{hg}\right)^2\exp\left[-\frac{\pi m c^2}{hg}\left(\tau-g\right)^2\right] \tag{5-30}$$

and

$$p\left(\overline{x}_2,s\,|\,\overline{x}_0,s_0\right)=\left(\frac{m}{hg}\right)^2\exp\left[-\frac{\pi m c^2}{hg}\left(g^2+\tau^2\right)\right] \tag{5-31}$$

The ratio of those two values is then:

$$\frac{p\left(\overline{x}_2,s|\,\overline{x}_0,s_0\right)}{p\left(\overline{x}_1,s|\,\overline{x}_0,s_0\right)}=\exp\left(-\frac{2\pi mc^2}{h}\tau\right) \qquad (5\text{-}32)$$

This value is always less than 1 and it even converges very rapidly to 0 when $\tau$ increases. For example, we have in the case of an electron:

$$\frac{2\,\pi mc^2}{h}=7.7634\ \text{x}10^{20}\ \text{s}^{-1} \qquad (5\text{-}33)$$

We normally observe a particle over a period of time very large compared to the reciprocal of this value and it thus appears to move (almost) always in agreement with relativity because the probability density for a spacelike displacement is negligibly small compared to the probability density for a timelike displacement. However, those two probability densities are nearly equal for very small values of $\tau$. The expression:

$$g_{ij}w^{ij}\left(s-s_0\right)=E\left\{\,g_{ij}\left(x^i-x_0^i\right)\left(x^j-x_0^j\right)|\,\overline{x}_0\right\} \qquad (5\text{-}34)$$

certainly incorporates displacements not allowed to a classical particle by relativity, that is displacements for which the quadratic expression $g_{ij}\left(x^i-x_0^i\right)\left(x^j-x_0^j\right)$ is not positive, and can thus admit apparently impossible values.

## 6. Non-relativistic limit

Coming back to the formula (5-25) we can prove that the limit of its first factor when c converges to $\infty$ is the Dirac measure $\delta\left[c\left(t-t_0-g+g_0\right)\right]$ and we can see that the second factor is nothing else than the solution of the corresponding non-relativistic problem if we identify $g-g_0$ with $t-t_0$. This is in complete agreement with a hypothesis that explains how the results of the non-relativistic stochastic theory must coincide with the limits of the corresponding results of the relativistic stochastic theory when the speed of light c is supposed to increase indefinitely to $\infty$. This must be done carefully however by taking care of two requisites.

The first one is so to say of a typographical nature and is imposed to us in order to make a clear notational distinction between the relativistic and non-relativistic random variables and functions thereof. This is why we shall designate by $\hat{p}\left(\hat{x},t\right)$ and $\hat{p}\left(\hat{x},t|\,\hat{x}_0,t_0\right)$ the normal and conditional probability densities, by $\tilde{v}_{\pm}^i\left(\hat{x},t\right)$ the drift vectors, by $\hat{w}_{\pm}^{ij}\left(\hat{x},t\right)$ the diffusion tensors and by $d\hat{x}$ the volume element in the three-dimensional Cartesian space of the variables $\hat{x}$. Note that we have:

$$d\overline{x}=c\,dt\,d\hat{x} \qquad (6\text{-}1)$$

The second requisite is related to the fact that the relativistic theory can make a choice among a wide family of dynamical parameters s, while the non-relativistic theory is restricted to use the classical time t as the only possible dynamical parameter. Remember that if s and S are two possible relativistic dynamical parameters, the functions $g\left(s\right)$ and $G\left(S\right)$ introduced in the first paragraph are actually identical. The limit process on s or S will thus lead to a unique result if we choose any of those functions as the dynamical parameter for the stochastic variables $\hat{x}^i$. As a consequence, we shall use interchangeably the expression $g(s)$ or the symbol t to designate the time in the non-relativistic notations that we have proposed.

We want now to grapple with the problem of describing the relation that must exist between the relativistic and non-relativistic stochastic quantum theories. This question is logically and almost obviously answered by requiring that we must recover a non-relativistic diffusion stochastic process characterized by a probability density $\hat{p}(\hat{x},t)$ and a conditional probability density $\hat{p}(\hat{x},t|\hat{x}_0,t_0)$ when we apply the non-relativistic limit $\lim c=\infty$ to a relativistic diffusion stochastic process characterized by a probability density $p(\overline{x})$ and a conditional probability density $p(\overline{x},s|\overline{x}_0,s_0)$. We shall therefore proceed in an inductive way in that direction by showing that the conditions:

$$\begin{cases} \lim\limits_{c=\infty} p(\overline{x}) = \hat{p}(\hat{x},g)\delta\left[c(t-g)\right] \\ \lim\limits_{c=\infty} p(\overline{x},s|\overline{x}_0,s_0) = \hat{p}(\hat{x},g|\hat{x}_0,g_0)\delta\left[c(t-t_0-g+g_0)\right] \end{cases} \tag{6-2}$$

with the abbreviations:

$$\begin{cases} g = g(s) \\ g_0 = g(s_0) \end{cases} \tag{6-3}$$

provide not only a very interesting mathematical hypothesis, but also probably the only physically acceptable thesis, because they guarantee the coherency between the relativistic and non-relativistic domains not only of the results obtained up to now, but also of the definitions introduced previously. We shall call this requisite the (relativistic) correspondence principle in spite of the fact that it could be the source of a potential confusion.

In the calculations to come, we shall not pay any attention to the ad hoc hypothesis that would allow us to exchange freely the limit and integration operations. The manipulation of δ functions under the integral sign will be done here in a very formal way, but you must be confident that the theory of the measures can prove them in a rigorous mathematical context.

We shall first investigate how the relativistic drift vector $v_{\pm}^{i}(\overline{x})$ and the non-relativistic drift vector $\hat{v}_{\pm}^{i}(\hat{x},s)$ are related to each other. When $i=0$ we can write:

$$\begin{aligned} \lim_{c=\infty} v_{\pm}^{0}(\overline{x}_0)\dot{g}(s_0)(s-s_0) &= \lim_{c=\infty} v^{0}(\overline{x}_0,s_0)(s-s_0) \\ &= \lim_{c=\infty}\int p(\overline{x},s|\overline{x}_0,s_0)\left[x^0(s)-x^0(s_0)\right]d\overline{x} \\ &= \int \hat{p}(\hat{x},g|\hat{x}_0,g_0)d\hat{x}\int c(t-t_0)\delta\left[c(t-t_0-g+g_0)\right]c\,dt \\ &= c(g-g_0) \\ &= c\,\dot{g}(s_0)(s-s_0) \end{aligned} \tag{6-4}$$

When $i \neq 0$ we must write instead:

$$
\begin{aligned}
\lim_{c=\infty} v^{i}_{\pm}\left(\overline{x}_0\right)\dot{g}\left(s_0\right)\left(s-s_0\right) &= \lim_{c=\infty}\int p\left(\overline{x},s\,|\,\overline{x}_0,s_0\right)\left[x^{i}\left(s\right)-x^{i}\left(s_0\right)\right]d\overline{x} \\
&= \int \hat{p}\left(\hat{x},g\,|\,\hat{x}_0,g_0\right)\left[\hat{x}^{i}\left(g\right)-\hat{x}^{i}\left(g_0\right)\right]d\hat{x}\int \delta\left[c\left(t-t_0-g+g_0\right)\right]c\,dt \\
&= \int \hat{p}\left(\hat{x},g\,|\,\hat{x}_0,g_0\right)\left[\hat{x}^{i}\left(g\right)-\hat{x}^{i}\left(g_0\right)\right]d\hat{x} \\
&= \hat{v}^{i}_{\pm}\left(\hat{x}_0,g_0\right)\left(g-g_0\right) \\
&= \hat{v}^{i}_{\pm}\left(\hat{x}_0,g_0\right)\dot{g}\left(s_0\right)\left(s-s_0\right)
\end{aligned}
\tag{6-5}
$$

We so obtain:

$$
\begin{cases}
\lim\limits_{c=\infty} v^{0}_{\pm}\left(\overline{x}\right)=c \\
\lim\limits_{c=\infty} v^{i}_{\pm}\left(\overline{x}\right)=\hat{v}^{i}_{\pm}\left(\hat{x},g\right) \qquad \text{for } i\neq 0
\end{cases}
\tag{6-6}
$$

These relations immediately imply that we have:

$$
\begin{cases}
\lim\limits_{c=\infty} v^{0}\left(\overline{x}\right)=c \\
\lim\limits_{c=\infty} v^{i}\left(\overline{x}\right)=\hat{v}^{i}\left(\hat{x},g\right) \qquad \text{for } i\neq 0
\end{cases}
\tag{6-7}
$$

We shall now investigate how the relativistic diffusion tensor $w^{ij}\left(\overline{x}\right)$ and the non-relativistic diffusion tensor $\hat{w}^{ij}\left(\hat{x},g\right)$ are related to each other. When $i\neq 0\neq j$ we can repeat the previous calculations that now leads to:

$$
\lim_{c=\infty} w^{ij}\left(\overline{x}_0\right)\dot{g}\left(s_0\right)\left(s-s_0\right)=\hat{w}^{ij}\left(\hat{x}_0,g_0\right)\dot{g}\left(s_0\right)\left(s-s_0\right)
\tag{6-8}
$$

We so obtain:

$$
\lim_{c=\infty} w^{ij}\left(\overline{x}\right)=\hat{w}^{ij}\left(\tilde{x},g\right) \qquad \text{for } \; i\neq 0\neq j
\tag{6-9}
$$

The relations (6-7) and (6-9) are not mere tautologies but must rather be interpreted as proving that the presented properties of the non-relativistic drift vectors $\hat{v}^{i}_{\pm}\left(\hat{x},g\right)$ and diffusion tensor $\hat{w}^{ij}\left(\hat{x},g\right)$ as being the non-relativistic limits of the corresponding relativistic drift vectors $v^{i}_{\pm}\left(\overline{x}\right)$ and diffusion tensor $w^{ij}\left(\overline{x}\right)$ are completely equivalent to their definitions based on the conditional expectations computed by using the non-relativistic conditional probability density $\hat{p}\left(\hat{x},g\,|\,\hat{x}_0,g_0\right)$ as this must be the case for a diffusion process.

We shall show that the limits (6-9) can be obtained in another way. Let us present this thesis more precisely. On the one hand, the non-relativistic diffusion process that is at the base of the non-relativistic stochastic quantum theory was supposed to be characterized by the diffusion tensor:

$$
\hat{w}^{ij}=\frac{h}{4\pi m}\hat{g}^{ij}
\tag{6-10}
$$

where $\hat{g}^{ij}$ is the metric tensor of the three-dimensional space of the variables $\hat{x}^{i}$. In a previous paragraph on the other hand, we have proposed for the relativistic diffusion process at the base of the relativistic stochastic quantum theory a diffusion tensor $w^{ij}(\bar{x})$ equal to:

$$w^{ij}(\bar{x}) = \frac{h}{4\pi m}\left[2\frac{v^i(\bar{x})v^j(\bar{x})}{|v(\bar{x})|^2} - g^{ij}\right] \tag{6-11}$$

We know that the relations that exist between the metric tensors $g^{ij}$ and $\tilde{g}^{ij}$ are:

$$\tilde{g}^{ij} = -g^{ij} \quad \text{for } i \neq 0 \neq j \tag{6-12}$$

Taking into account the relations (6-7), we obtain:

$$\begin{aligned}\lim_{c=\infty}|v(\bar{x})|^2 &= \lim_{c=\infty} g_{ij}v^i(\bar{x})v^i(\bar{x}) \\ &= \lim_{c=\infty}\left(c^2 - \hat{g}_{ij}\hat{v}^i(\hat{x},t)\hat{v}^i(\hat{x},t)\right) \\ &= c^2\end{aligned} \tag{6-13}$$

that by the way is not in contradiction with the normalization condition (**Error! Reference source not found.**). The results:

$$\begin{cases}\lim\limits_{c=\infty} v^0(\bar{x})/|v(\bar{x})| = 1 \\ \lim\limits_{c=\infty} v^i(\bar{x})/|v(\bar{x})| = 0 \text{ for } i \neq 0\end{cases} \tag{6-14}$$

that can be deduced there from lead to the limit:

$$\begin{aligned}\lim_{c=\infty} w^{ij} &= \lim_{c=\infty}\frac{h}{4\pi m}\left[2\frac{v^i(\bar{x})v^j(\bar{x})}{|v(\bar{x})|^2} - g^{ij}\right] \\ &= \frac{h}{4\pi m}\hat{g}^{ij} \qquad\qquad \text{for } i \neq 0 \neq j \\ &= \tilde{w}^{ij}\end{aligned} \tag{6-15}$$

in complete agreement with (6-9). At the same limit, we should obtain for the component $w^{00}$:

$$\begin{aligned}\lim_{c=\infty} w^{00}(\bar{x}) &= \lim_{c=\infty}\frac{h}{4\pi m}\left[2\frac{v^0(\bar{x})v^0(\bar{x})}{c^2} - g^{00}\right] \\ &= \frac{h}{4\pi m}\end{aligned} \tag{6-16}$$

and this would place the variable ct on an equal footing with the variables x, y and z.

To conclude this paragraph we shall simply mention that the correspondence principle applies equally well to the normalization properties of the normal and conditional probability densities and also to the Chapman, Fokker-Planck and Kolmogorov equations.

## 7. General theory of the relativistic systems.

We have seen that the relativistic Fokker-Planck and Kolmogorov equations can be written by using exclusively the drift vectors $v^i_{\pm}(\overline{x})$ and the diffusion tensor $w^{ij}(\overline{x})$ that do not depend on the dynamical parameter s. We can then expect that many of the results obtained for the non-relativistic stationary diffusion processes can be transposed in the present context word for word so to say.

Here after we shall study the Kolmogorov equations (5-6) and (5-7) concerning the conditional probability density $p(\overline{x}_1, s_1 \mid \overline{x}_0, s_0)$. We shall restrict our study of those equations to the forward case where we have $s > s_0$ because all that can be said about them is also valid for the reverse case with due changes.

Those equations can be solved by the method of separation of the variables. Remember that this amounts to represent the general solutions of the equations (5-6) and (5-7) by sums of particular solutions of the form $X^0(\overline{x}_0)S^0(s_0)$ and $X(\overline{x})S(s)$ respectively. If we introduce those particular solutions into their corresponding equations and if we divide by those solutions, we arrive at the equations:

$$-\frac{1}{\dot{g}(s_0)}\frac{\dot{S}^0(s_0)}{S^0(s_0)} = \frac{1}{X^0(\overline{x}_0)}\left[ w^{ij}(\overline{x}_0)\,\frac{\partial^2 X^0(\overline{x}_0)}{\partial x^i_0\,\partial x^j_0} + v^i_+(\overline{x}_0)\,\frac{\partial X^0(\overline{x}_0)}{\partial x^i_0}\right] \tag{7-1}$$

and:

$$\frac{1}{\dot{g}(s)}\frac{\dot{S}(s)}{S(s)} = \frac{1}{X(\overline{x})}\left[\frac{\partial^2\left[ w^{ij}(\overline{x})X(\overline{x})\right]}{\partial x^i\,\partial x^j} - \frac{\partial\left[ v^i_+(\overline{x})X(\overline{x})\right]}{\partial x^i}\right] \tag{7-2}$$

The left hand side member of equation (7-1) does not depend on the variable $\overline{x}_0$ while the right hand side member does not depend on the variable $s_0$ so that their common value must be a constant that we shall designate by $-\lambda$. Similarly, the common value of the two members of equation (7-2) must also be a constant that we shall designate by $-\mu$. Then equations for $S^0(s_0)$ and $S(s)$ can be written in the forms:

$$\frac{\dot{S}^0(s_0)}{S^0(s_0)} = \lambda\,\dot{g}(s_0) \tag{7-3}$$

$$\frac{\dot{S}(s)}{S(s)} = -\mu\,\dot{g}(s) \tag{7-4}$$

and can be solved immediately to give:

$$S^0(s_0) = \exp\left[\lambda\, g(s_0)\right] \tag{7-5}$$

$$S(s) = \exp\left[-\mu g(s)\right] \tag{7-6}$$

while the equations for $X^0(\overline{x}_0)$ and $X(\overline{x})$ take the forms:

$$w^{ij}(\bar{x}_0)\ \frac{\partial^2 X^0(\bar{x}_0)}{\partial x_0^i\ \partial x_0^j}+v_+^i(\bar{x}_0)\ \frac{\partial X^0(\bar{x}_0)}{\partial x_0^i}+\lambda X^0(\bar{x}_0)=0 \qquad (7\text{-}7)$$

and:

$$\frac{\partial^2}{\partial x^i\ \partial x^j}\left[\ w^{ij}(\bar{x})\ X(\bar{x})\ \right]-\frac{\partial}{\partial x^i}\left[\ v_+^i(\bar{x})\ X(\bar{x})\ \right]+\mu X(\bar{x})=0 \qquad (7\text{-}8)$$

We know that these equations have non identically zero solutions called the eigenfunctions only for a discrete denumerably infinite set of values for the parameters $\lambda$ and $\mu$ called the eigenvalues. Let us designate by $X_i^0(\bar{x}_0)$ the eigenfunction of the equation (7-7) associated with the eigenvalue $\lambda_i$ and let us similarly designate by $X_i(\bar{x})$ the eigenfunction of the equation (7-8) associated with the eigenvalue $\mu_i$. We know that those solutions can be normalized and indexed by non-negative integer numbers in such a way that we have:

$$\lambda_i=\mu_i \qquad (7\text{-}9)$$

and:

$$\int X_i(\bar{x})X_k^0(\bar{x})d\bar{x}=\delta_{ik} \qquad (7\text{-}10)$$

with:

$$\delta_{ik}=\begin{cases}0 & \text{if } i\neq k\\ 1 & \text{if } i=k\end{cases} \qquad (7\text{-}11)$$

Then the solution $p(\bar{x},s\,|\,\bar{x}_0,s_0)$ of the Kolmogorov equations verifying the initial condition:

$$p(\bar{x},s_0\,|\,\bar{x}_0,s_0)=\delta(\bar{x}-\bar{x}_0) \qquad (7\text{-}12)$$

is given by:

$$p(\bar{x},s\,|\,\bar{x}_0,s_0)=\sum_{n=0}^{\infty}X_n(\bar{x})X_n^0(\bar{x}_0)\exp\left\{-\lambda_n\left[g(s)-g(s_0)\right]\right\} \qquad (7\text{-}13)$$

The condition $v_i=0$ simplifies considerably the theory of the Kolmogorov equations because they are then equivalent to a self-adjoint equation that is known to possess real eigenvalues and eigenfunctions. The bad news is that this condition is equivalent to an intolerable restrictive physical condition. We have seen indeed as a result of our variational principle that the vector $m v_i + e A_i$ must be a gradient. Thus, the electromagnetic potential $A_i$ must also be a gradient if $v_i=0$ and the electromagnetic field $B_{ij}$ given by:

$$B_{ij}=\partial_j A_i-\partial_i A_j \qquad (7\text{-}14)$$

must inevitably be equal to zero. Conversely, if the electromagnetic field $B_{ij}$ is different from zero, the vector $v_i$ must also be different from zero. If not, the electromagnetic field specified in the right hand side member of the second Nelson equation would play no role at all. So except in the case of the free particle the equations (7-7) and (7-8) are no longer equivalent to a self-adjoint equation and their non-zero eigenvalues $\lambda_i = \mu_i$ and associated eigenfunctions $X_i^0(\overline{x})$ and $X_i^1(\overline{x})$ are now possibly complex. In that case and as proposed before, we shall modify the notations by letting the index i run from $-\infty$ to $+\infty$ with the conditions:

$$\begin{cases} \lambda_{-i} = \lambda_i^* \\ X_{-i}^0(\overline{x}_0) = X_i^0(\overline{x}_0)^* \\ X_{-i}^1(\overline{x}_1) = X_i^1(\overline{x}_1)^* \end{cases} \tag{7-15}$$

and to modify accordingly equation (7-13) by writing it in the form:

$$p(\overline{x},s|\,\overline{x}_0,s_0) = \sum_{n=-\infty}^{\infty} X_n(\overline{x}) X_n^0(\overline{x}_0) \exp\left\{-\lambda_n\left[g(s)-g(s_0)\right]\right\} \tag{7-16}$$

The general expression obtained for the conditional probability density $p(\overline{x},s \mid \overline{x}_0,s_0)$ shows that it depends on s and $s_0$ by the difference $g(s)-g(s_0)$ only. Because the diffusion process $\overline{x}(t)$ is a Markov process, we can write:

$$\begin{aligned} p(\overline{x}_1,s_1;\ \cdots\ ;\overline{x}_n,s_n) &= \frac{p(\overline{x}_1,s_1;\ \cdots\ ;\overline{x}_n,s_n)}{p(\overline{x}_1,s_1;\ \cdots\ ;\overline{x}_{n-1},s_{n-1})} \frac{p(\overline{x}_1,s_1;\ \cdots\ ;\overline{x}_{n-1},s_{n-1})}{p(\overline{x}_1,s_1;\ \cdots\ ;\overline{x}_{n-2},s_{n-2})} \cdots \\ &\qquad \cdots \frac{p(\overline{x}_1,s_1;\overline{x}_2,s_2)}{p(\overline{x}_1,s_1)} p(\overline{x}_1,s_1) \\ &= p(\overline{x}_n,s_n \mid \overline{x}_1,s_1;\cdots;\overline{x}_{n-1},s_{n-1}) p(\overline{x}_{n-1},s_{n-1} \mid \overline{x}_1,s_1;\cdots;\overline{x}_{n-2},s_{n-2}) \cdots \\ &\qquad \cdots p(\overline{x}_2,s_2 \mid \overline{x}_1,s_1) p(\overline{x}_1,s_1) \\ &= p(\overline{x}_n,s_n \mid \overline{x}_{n-1},s_{n-1}) p(\overline{x}_{n-1},s_{n-1} \mid \overline{x}_{n-2},s_{n-2}) \cdots p(\overline{x}_2,s_2 \mid \overline{x}_1,s_1) p(\overline{x}_1) \end{aligned} \tag{7-17}$$

Taking into account the previous remark this formula proves that $p(\overline{x}_1,s_1;\ \cdots\ ;\overline{x}_n,s_n)$ depends on the variables $s_1,s_2,\cdots,s_n$ by the differences $g(s_2)-g(s_1)$, ..., $g(s_n)-g(s_{n-1})$ only.

It is obvious that the constant function $X_0^0(\overline{x}) = 1$ is a solution of the equation (7-7) associated with the eigenvalue $\lambda = 0$ and that according to the Fokker-Planck equation, the function $X_0(\overline{x}) = p(\overline{x})$ is a solution of the equation (7-8) associated with the same eigenvalue $\mu = 0$. We have for this pair:

$$\begin{aligned} \int X_0^0(\overline{x}) X_0^1(\overline{x}) d\overline{x} &= \int p(\overline{x}) d\overline{x} \\ &= 1 \end{aligned} \tag{7-18}$$

and we have not to worry about multiplicative constants for this pair of solutions.

It is obvious that if (the real part of) all the non-zero eigenvalues are positive, we can write:

$$\lim_{g(s)=\infty} p\left(\overline{x},s\,|\,\overline{x}_0,s_0\right) = \lim_{g(s)=\infty} \sum_n X_n\left(\overline{x}\right) X_n^0\left(\overline{x}_0\right) \exp\left\{-\lambda_n\left[g(s)-g(s_0)\right]\right\}$$
$$= X_0\left(\overline{x}\right) X_0^0\left(\overline{x}_0\right) \qquad (7\text{-}19)$$
$$= p\left(\overline{x}\right)$$

and this clearly shows that the physical phenomenon described by the Kolmogorov equations has an inherent tendency to go or to return to the stable state characterised by the probability density $p\left(\overline{x}\right)$. The H-theorem proved in the previous paragraph shows that this must be the case.

Our intuition leads us to think that the conjugate solutions must have something in common. In the case of the hydrogen atom for example, we can claim using the parlance of the orthodox theory that the pair of the eigenfunctions $X_0^0\left(\overline{x}_0\right)=1$ and $X_0\left(\overline{x}\right)=p\left(\overline{x}\right)$ associated with the eigenvalues $\lambda_0=\mu_0=0$ corresponds to the ground state and we can imagine that all other pairs $X_i^0\left(\overline{x}\right)$ and $X_i\left(\overline{x}\right)$ associated with the eigenvalues $\lambda_i=\mu_i\neq 0$ describe states having some physical characteristics in common (like the energy?) and possessing some other physical characteristics in their own right (like the spin?). It is all the same curious that the spin was introduced in the Dirac theory because the Klein-Gordon equation is unable to treat correctly even the simple case of the atom of hydrogen where an electric field alone is present.

It would be welcome to confirm those ideas by studying a very simple example, but we can anticipate that it will be nonetheless a difficult task because complete separability of the equations (7-7) and (7-8) is no longer possible in the case where $v_i\neq 0$. This operation amounts indeed to replace a second order differential equation in m variables $x^i$ by an equivalent system of m second order differential equations in only one variable $x^i$ at a time. It is always possible to associate a self-adjoint differential equation to those partial equations and the same conclusion must also be valid for the complete differential equation, calling for $v_i=0$.

## 8. Conclusion

Several authors have proposed their own definition of a relativistic diffusion process, but they do not pay enough attention (if any) to its theory, including the Kolmogorov and Fokker-Planck equations, the stochastic differential and integral equations, the reversibility properties, the variational principle and all the paraphernalia, that we have used. This is because they aim primarily at rediscovering the Klein-Gordon equation, what in our opinion is not a good idea. Conversely, other authors pay too much attention to a supposedly universally valid theorem due to Hakim, stating that a relativistic Markov process cannot represent a diffusion process because it is always characterized by $w^{ij}=0$, so that they feel obliged to introduce rather special hypotheses in order to develop their idea anyhow. However, this negative result is obtained when studying a relativistic Markov process that describes the evolution of a stochastic vector in a 7 dimensional space identical with the product of the 4 dimensional Minkowski space and of a 3 dimensional velocity space obtained by reducing the complete 4 dimensional velocity space according to the relativistic condition $\overline{v}^2=c^2$

The present theory is obviously very different, but is it better? As a first favourable argument, we must recognize that it clearly displays a remarkable internal consistency despite the fact that its basic hypotheses are sometimes bizarre and questionable. In addition, several welcome results concerning the non-relativistic theory are still valid in the relativistic domain, simply because they are based on general properties of a Markov process. Let us quote as examples that:

- Uncertainty relations exist even for the time stochastic variable,
- The trajectories of the particles are continuous functions of the dynamical parameter s, but possess nowhere a derivative with respect to it,
- The collapse of the wave function is a mere myth and is simply replaced by an updating of the initial conditions. This eliminates by the same token the much discussed measurement problem.
- The H theorem can be invoked to prove that any state of a system finally converges to the fundamental state characterized by the normal probability density,
- The ergodic theorem can be invoked to replace mean values with respect to the dynamical parameter s by mathematical conditional expectations.

This paper however waits for a continuation. So for example, we have not been able to solve the apparently simple problem of the Coulomb field. So decisive arguments based on a comparison of the energy levels and on the automatic intrusion of the spin are unfortunately missing at this time,